\documentclass[11pt,a4paper]{article}
\pdfoutput=1
\usepackage{jheppub}

\usepackage{ifpdf}

\usepackage{afterpage}

\usepackage{amssymb}
\usepackage{amsmath}
\usepackage{graphicx}
\usepackage{longtable}
\usepackage{verbatim}
\usepackage{amsfonts}
\usepackage{xcolor}
\usepackage{ulem}

\newcommand{\bear}{\begin{array}}
\newcommand{\ear}{\end{array}}

\newcommand{\beq}{\begin{eqnarray}}
\newcommand{\eeq}{\end{eqnarray}}
\newcommand{\beqa}{\begin{eqnarray}}
\newcommand{\eeqa}{\end{eqnarray}}

\def\OMIT#1{{}}
\newcommand{\lsim}{\mathrel{\rlap{\lower4pt\hbox{\hskip1pt$\sim$}}
    \raise1pt\hbox{$<$}}}         
\newcommand{\gsim}{\mathrel{\rlap{\lower4pt\hbox{\hskip1pt$\sim$}}
    \raise1pt\hbox{$>$}}}         

\newcommand{\be}{\begin{equation}}
\newcommand{\ee}{\end{equation}}
\newcommand{\ba}{\begin{eqnarray}}
\newcommand{\ea}{\end{eqnarray}}

\def\lsim{\mathrel{\rlap{\lower4pt\hbox{\hskip1pt$\sim$}}
    \raise1pt\hbox{$<$}}}         
\def\gsim{\mathrel{\rlap{\lower4pt\hbox{\hskip1pt$\sim$}}
    \raise1pt\hbox{$>$}}}         

\newcommand{\iu}{{\mathrm i}}
\newcommand{\E}{{\mathrm e}}

\newcommand{\tF}{{\widetilde F}}
\newcommand{\tG}{{\widetilde G}}
\newcommand{\tH}{{\widetilde H}}
\newcommand{\tQ}{{\widetilde Q}}
\newcommand{\tL}{{\widetilde L}}
\newcommand{\tN}{{\widetilde N}}

\renewcommand{\tilde}{\widetilde}

\vspace*{-30mm}

\title{\boldmath Experimental Targets for Photon Couplings of the QCD Axion}
\author[a]{Prateek Agrawal,}
\author[b]{JiJi Fan,} 
\author[a]{Matthew Reece}
\author[c]{and Lian-Tao Wang}
\affiliation[a]{Department of Physics, Harvard University, Cambridge, MA 02138, USA}
\affiliation[b]{Department of Physics, Brown University, Providence, RI, 02912, USA}
\affiliation[c]{Enrico Fermi Institute and Kavli Institute for Cosmological Physics, \\University of Chicago, Chicago, IL 60637, USA}

\vspace*{1cm}

\abstract{The QCD axion's coupling to photons is often assumed to lie
in a narrow band as a function of the axion mass. We demonstrate that
several simple mechanisms, in addition to the photophilic clockwork
axion already in the literature, can significantly extend the allowed
range of couplings. Some mechanisms we present generalize the KNP
alignment scenario, widely studied as a model of inflation, to the
phenomenology of a QCD axion. In particular we present
KSVZ-like
realizations of two-axion KNP alignment and of the clockwork
mechanism. Such a ``confinement tower'' realization of clockwork may
prove useful in a variety of model-building contexts. We also show
that kinetic mixing of the QCD axion with a lighter axion-like
particle can dramatically alter the QCD axion's coupling to photons,
differing from the other models we present by allowing non-quantized
couplings. The simple models that we present fully cover the range of
axion--photon couplings that could be probed by experiments. They
motivate growing axion detection efforts over a wide space of masses
and couplings. }
\begin{document}
\maketitle

\section{Introduction}

The QCD axion is the most appealing simple solution to the strong CP
problem~\cite{Peccei:1977ur, Peccei:1977hh, Wilczek:1977pj,
Weinberg:1977ma, Kim:1979if, Shifman:1979if, Zhitnitsky:1980tq,
Dine:1981rt} as well as a classic dark matter
benchmark~\cite{Preskill:1982cy, Dine:1982ah, Abbott:1982af}. Given
its very weak coupling to the standard model, searches to discover it
have proved to be challenging. Yet experimental efforts have been
growing very rapidly recently~\cite{Asztalos:2009yp, Graham:2013gfa,
Armengaud:2014gea, Horns:2012jf, Budker:2013hfa, Arvanitaki:2014dfa,
Kahn:2016aff} with several of them aiming at detecting axion--photon
couplings. It is thus important to chart the motivated parameter space
for this coupling.

An axion is a periodic field, $a \cong a + 2\pi F_a$. This constrains
its couplings to gauge fields, as $\theta$ has period $2\pi$ in a
coupling $\frac{\theta}{32 \pi^2} e^2 F_{\mu \nu}\tF^{\mu \nu}$, where
the dual gauge field $\tilde{F}_{\mu\nu} = \frac{1}{2}
\epsilon_{\mu\nu\rho\sigma}F^{\rho\sigma}$. (Recall that even for a
$U(1)$ gauge theory, the $\theta$ term is physical, as manifested in
the Witten effect \cite{Witten:1979ey}.) Compatibility of the axion
period and the $\theta$ angle period requires that when we have a
coupling of an axion to gauge fields (abelian or nonabelian) of the
form \be
k \frac{\alpha}{8 \pi} \frac{a}{F_a} F_{\mu \nu} \tF^{\mu \nu}, \label{eq:quantizedcoupling}
\ee 
with the gauge field canonically normalized, $\alpha = e^2/(4\pi)$, and $e$ the coupling to a minimum-charge particle, the prefactor $k$ must be an integer.\footnote{Because we normalize Standard Model charges so that the smallest is $1/3$ rather than $1$, the quantization of the coefficient is in units of $1/9$ rather than $1$. The periodicity of the theta angle can be altered by integer factors when the gauge group is quotiented by a discrete subgroup \cite{Witten:1980sp, Witten:1998uka}, and even in the Standard Model this leads to some ambiguity regarding the proper periodicity of the QED $\theta$ angle \cite{Tong:2017oea}.}

The QCD axion's mass is determined by nonperturbative dynamics resulting from its coupling to gluons,
\be
N \frac{\alpha_s}{8\pi} \frac{a}{F_a} G^a_{\mu \nu} \tG^{a\mu\nu} = \frac{\alpha_s}{8\pi} \frac{a}{f_a} G^a_{\mu \nu} \tG^{a\mu\nu}.   \label{eq:QCDcouplingdef}
\ee
Here $N$ is an integer and we have defined the effective decay constant
\be
f_a \equiv F_a/N.
\ee
In this paper our focus will be on the axion--photon coupling,
$-\frac{g_{a\gamma\gamma}}{4} a F \tilde{F}$. This is a sum of two
contributions: the IR one from mixing between axion and QCD mesons \cite{Kaplan:1985dv, Srednicki:1985xd, Georgi:1986df, Svrcek:2006yi}, with
\beq
g_{a\gamma\gamma}^{\rm IR} =  - 1.92(4) \frac{\alpha_{\rm em}}{2\pi f_a}, 
\label{eq:axionpionmixing}
\eeq
where $\alpha_{\rm em}$ is the electromagnetic coupling strength and $f_a$ is the effective decay constant introduced above. The number (4) indicates the NLO correction~\cite{diCortona:2015ldu}. The UV contribution to the axion--photon coupling is model-dependent. It usually takes the form 
\beq
g_{a\gamma\gamma}^{\rm UV} =r \frac{\alpha_{\rm em}}{2\pi f_a}, \quad {\rm with} \quad r=\frac{E}{N}
\eeq
where $E$ and $N$ are the (discrete) electromagnetic and QCD anomaly coefficients
of the PQ symmetry respectively. 
The IR contribution indicates the smallest size of the axion--photon
coupling, provided that there is no accidental cancelation between the
UV and IR contributions. In models where
$E/N = 2$,\footnote{This could be realized in specific models, e.g., a
KSVZ model with one set of vector-like heavy quarks, which is a color
fundamental and electroweak singlet, and one set of vector-like heavy
leptons which only carry hypercharge $1 (-1)$.}
the axion--photon coupling
is reduced
by a factor of $\sim 20$~\cite{Giudice:2012zp}. More
  extreme
  tuning is possible by considering multiple representations or
through a kinetic mixing contribution. Notice that mixing of multiple axions can appear to evade the quantization rule (\ref{eq:quantizedcoupling}), because the kinetic and mass terms may not be diagonal in a basis where the axion shift symmetries are diagonal. For clarity and pedagogical completeness, we elaborate on the origin of the non-quantized coupling (\ref{eq:axionpionmixing}) in Appendix \ref{massmix}.

The question is then: what is the upper bound of the QCD axion--photon
coupling theoretically? Traditionally it is assumed that UV and IR
contributions are of the same order and $g_{a\gamma\gamma} \sim {\cal
O}(1) \alpha_{\rm em}/(2\pi f_a)$. 
A variety of specific models realizing different ${\cal O}(1)$ coefficients have been used to define a standard band that is often plotted \cite{Kim:1998va}. 
More thorough recent analyses
demonstrate that in the standard KSVZ \cite{Kim:1979if, Shifman:1979if} and DFSZ \cite{Zhitnitsky:1980tq, Dine:1981rt} frameworks, it is true
that $g_{a\gamma\gamma} \sim {\cal O}(1) \alpha_{\rm em}/(2\pi f_a)$
for most representations of heavy matter charged under the PQ symmetry
and the standard model gauge groups~\cite{DiLuzio:2016sbl,
DiLuzio:2017pfr}. Yet special representations of KSVZ fermions and
their combinations or multiple Higgses (9
Higgses) with particular PQ charges in the DFSZ model could give rise to
larger couplings~\cite{DiLuzio:2016sbl, DiLuzio:2017pfr}. In this
case, requiring that the Landau poles of the SM gauge couplings are
above the Planck scale in the presence of the new matter charged under
the SM gauge group, the QCD axion's coupling to photons could be
maximally enhanced to 170/3 (KSVZ model) and 524/3 (DFSZ model).
Recently it has also been proposed that axion coupling to photons
could be enhanced exponentially in a clockwork axion
scenario~\cite{Farina:2016tgd} (based on the clockwork idea of
\cite{Choi:2015fiu, Kaplan:2015fuy}, which had precursors in
\cite{Dvali:2007hz, Choi:2014rja}). 
This clockwork photophilic axion relies on a particular structure of
multiple scalars. It serves as a very interesting proof of concept
that axion-photon couplings could be enhanced significantly.

In this article, we will showcase several different mechanisms that
can achieve a large axion-photon coupling. Our goal is not just to
prove that large couplings are possible, since the clockwork
photophilic axion model already demonstrates that; rather, we aim to
identify qualitatively different UV completions and explain the quantitative degree to which they
can enhance the axion-photon coupling. Our main point will be that
very simple extensions of familiar models of the QCD axion can lead to
a substantial enhancement of $g_{a\gamma\gamma}$. The mechanisms
include Kim-Nilles-Peloso (KNP) alignment \cite{Kim:2004rp} of two or
more axions and 
kinetic mixing of multiple axions. 
We realize the KNP alignment mechanism with hidden confining gauge
groups~\cite{Choi:2015fiu}. 
Models which iterate KNP alignment with a tower
of confining gauge groups provide a useful realization of the clockwork
mechanism. 
Among these mechanisms, the ones based on large PQ
charge or alignment predict quantized couplings while kinetic mixing
could give rise to non-quantized couplings. Together they motivate a much
broader experimental parameter space for the QCD axion.

As a byproduct, we note that the KSVZ-like constructions of KNP
alignment and clockwork models that we construct may be more generally
useful for phenomenology. For the most part, KNP alignment has been
discussed in the context of axions arising from extra-dimensional
gauge fields, while clockwork was based on theories of many scalars
with highly constrained quartic interactions.
An alternative is a KSVZ-like approach to KNP alignment. We present a simple nonsupersymmetric realization of this idea. A similar supersymmetric construction appeared in \cite{Choi:2015fiu}, while a prototype of our nonsupersymmetric approach appeared recently in \cite{Agrawal:2017eqm}. The version we present here
differs in relying entirely on choices of gauge representations rather
than a large number of fermion fields to obtain an enhancement. A rather different realization of clockwork based on a sequence of confining gauge groups also appeared recently in \cite{Coy:2017yex}; we will comment below on the similarities and differences to our approach.

In several mechanisms we present, new matter with standard model
hypercharge will accelerate the running of the $U(1)_Y$ gauge
coupling. While the requiring the Landau pole to be above the Planck
scale is not necessary, we will follow Ref.~\cite{DiLuzio:2016sbl,
DiLuzio:2017pfr} to adopt it as a theoretical constraint. 
We also restrict that all the fields in the model to be have no higher
than two-index representations of any non-Abelian gauge group. In
part, this is because all of the physics we are interested in can be
illustrated in simple models with only adjoint and fundamental
representations. A further motivation is that models with light matter
only in low-dimensional representations may be more UV completable. In
D-brane models, one finds only two-index representations because a
string has only two endpoints to attach to branes.  In the heterotic
string the story is more complicated, but similar statements are true
at low Kac-Moody levels (see e.g.~\S17.1 of \cite{Ibanez:2012zz}). In
short, we expect that by avoiding large charge assignments we obtain
easier compatibility with quantum gravity. 

We do not consider in detail using large hypercharges (or equivalently
a large number of fields with hypercharge) to boost the axion--photon
coupling. In addition to being exotic without any dynamical reason,
large hypercharges are also subject to the Landau pole constraint.
Requiring the Landau pole of $U(1)_Y$ to be above the Planck scale
$\sim 10^{18}$ GeV limits the
hypercharge of the heavy matter to be $\lesssim 6$, which leads to an
enhancement $\lesssim 100$.\footnote{The estimated number is obtained
  assuming a KSVZ model with one pair of vector-like quarks (without
  charge) and one pair of vector-like leptons which is only charged
  under $U(1)_Y$. This upper bound holds as long as the fermion mass
  is below $10^{17}$ GeV.}  A closely related possible method that may enhance the axion--photon coupling is to use large PQ charges. We will discuss it in appendix \ref{sec:model1} and demonstrate that due to constraint on the heavy fermions' mass, the enhancement is also limited to be below $32$. 

The same mechanisms that can be used to enlarge the QCD axion's coupling to photons could be used to enlarge the couplings to dark photons, which can help to make a wider range of QCD axion decay constants phenomenologically viable by altering the early universe cosmology \cite{Agrawal:2017eqm}. More generally, the idea of alignment (through charges as in KNP or through kinetic mixing) has played a major role in recent models of inflation, but a relatively limited role in other particle physics phenomenology (though see \cite{Higaki:2015jag, Higaki:2016yqk}). By illustrating simple renormalizable UV completions of alignment models, based on the same ideas as the original KSVZ axion model, we hope to spread these useful model-building tools to a wider phenomenological community.

\section{Scenario I: Alignment Mechanism} 
\label{sec:model2}
The KNP alignment mechanism has been proposed and studied extensively
in the natural inflation context~\cite{Kim:2004rp, Choi:2014rja,
Tye:2014tja, Kappl:2014lra, Ben-Dayan:2014zsa, Bai:2014coa,
delaFuente:2014aca}. In this section, we describe a KSVZ
completion of the KNP alignment. 

Let us first briefly review KNP alignment. In this scenario, we need at least two axion fields $a(x)$ and $b(x)$, both coupling to the gluons of a hidden gauge group $SU(M)_h$ and QCD gluons. The basic mechanism could be described by the Lagrangian
\beq
{\cal L} 
= 
&-& \frac{1}{4} \left(H_{\mu\nu} H^{\mu\nu} 
+ G_{\mu\nu} G^{\mu\nu} 
+ F_{\mu\nu} F^{\mu\nu}\right)  \nonumber\\
&+& \frac{\alpha_h}{8 \pi F_0}  (a+ M^\alpha b) H_{\mu\nu} \tilde{H}^{\mu\nu} 
+  \frac{\alpha_s}{8 \pi F_0}  b G_{\mu\nu} \tilde{G}^{\mu\nu} +  \frac{\alpha_{\rm em}}{8 \pi F_0}  M^\beta a F_{\mu\nu} \tilde{F}^{\mu\nu},
\eeq
where $H$ is the field strength of $SU(M)_h$. The powers of $M$ in the
anomaly coefficients, $\alpha, \beta \geq 1$, are some non-negative
integer powers depending on the particle content of the model. Note
that this Lagrangian is just illustrative. We have assumed that
$a$ and $b$ have the same period $F_0$ for simplicity. We highlight
the $M$ dependence and ignore ${\cal O}(1)$ numbers that could arise
in a full model. 

The heavy hidden gauge group confines at a scale $\Lambda_H \gg \Lambda_{\rm QCD}$ and leads to a heavy axion, which is a linear combination of $a$ and $b$ (mostly $b$). Effectively we can set 
\beq
a+ M^\alpha b = 0 \Rightarrow b = - M^{-\alpha} a
\eeq
to integrate out the heavy axion. In the low energy effective theory, we find the couplings of the light axion $a$ to be,
\beq
-\frac{\alpha_s}{8 \pi F_0} M^{-\alpha} a G \tilde{G} + \frac{\alpha_{\rm em}}{8 \pi F_0} M^\beta a F \tilde{F}.
\label{eq:align0}
\eeq
The first term suggests that the effective decay constant in this case is $f_a = M^\alpha F_0$, which could be significantly larger than the period $F_0$ in the UV theory.\footnote{In fact, because the light effective axion winds around the $(a,b)$ space, its period is also $f_a$ in the IR theory. We will treat these subtleties more carefully in the detailed example below.} Then the QCD axion coupling to the photons is then enhanced by $M^{\alpha+\beta}$:
\beq
r \equiv g_{a\gamma \gamma}^{\rm UV} \left(\frac{\alpha_{\rm em}}{8 \pi f_a }\right)^{-1}=  M^{\alpha+\beta}
\eeq
Below we will show a simple KSVZ type model with $\alpha = 1$ and
$\beta=2$ such that the enhancement scales as $M^3$. 

\subsection{A UV Completion Based on One Confining Hidden Gauge Group} 
\label{sec:model}

\begin{table}[h!]
\centering
	\begin{tabular}{| c  | c  c  c c c |}
	\hline
	 & $SU(M)_h$ & $SU(3)_C$ & $U(1)_Y$	& $U(1)_{PQ; 1}$ & $U(1)_{PQ; 2}$ \\ \hline \hline
	 $\phi_1$ & 1  & 1 & 0 & $-1$ & 0 \\ 
	 $Q_{1a} ({\widetilde Q}_{1a}) $ & Adj & 1 & 1 ($-1$) & 1(0) & 0 (0) \\ 
	 $Q_{1b} ({\widetilde Q}_{1b})$ & 1 & 3 ($\overline{3}$) & 0 & 1(0) & 0 (0)\\ 
	 $Q_{1c} ({\widetilde Q}_{1c})$ & $M$ ($\overline{M}$) & 1 & 0 & 1(0) & 0 (0) \\ \hline  \hline
	  $\phi_2$ & 1  & 1 & 0 & 0 & $-1$ \\ 
         $Q_{2a} ({\widetilde Q}_{2a}) $ & Adj & 1 & 0  & 0 & 1(0) \\  
	 $Q_{2b} ({\widetilde Q}_{2b})$ & 1 & 3 ($\overline{3}$) & 0 & 0 & 1(0) \\ \hline
	\end{tabular}
\caption{Particle content of a bi-axion alignment model. } \label{table:content}
\end{table}

The particle content is shown in Table~\ref{table:content}. The model
is a variant of the KSVZ model. $\phi_1, \phi_2$ are two
independent PQ fields associated with two $U(1)_{\rm PQ}$'s which we
will assume break at the same scale $F_0$ for simplicity. Below $F_0$,
there are two axion fields $a_1$ and $a_2$, which are the angular
degrees of freedom of $\phi_1$ and $\phi_2$ respectively. We take the
PQ charges of both $\phi_1$ and $\phi_2$ to be $-1$. In addition, we
have several sets of vector-like fermions. $Q$ and ${\widetilde Q}$ form a
vector-like pair. All the fermions with a subscript 1, $Q_1$'s, couple
to $\phi_1$ while $Q_2$'s couple to $\phi_2$, as implied by the PQ charge assignments: $y_1 \phi_1
Q_1{\widetilde Q}_1 + y_2 \phi_2 Q_2 {\widetilde Q}_2$. We also assume
all the heavy fermions are weak singlets. The key feature of this
model is that $\phi_1$ and $\phi_2$ couple to fermions with the same
representations of $SU(M)_h$ and $SU(3)_c$ to guarantee the alignment
of the heavy axion and QCD axion, except for one set of fermion
($Q_{1c}$ in the specific example), which only couples to one of the
PQ fields. 

For convenience, because we will make use of it extensively below, we quote here the axion--gauge field coupling generated by integrating out massive fermions. A mass term $m(\phi) Q \tilde{Q}$, with $m$ a general function of PQ-charged scalars and $Q$, $\tilde{Q}$ in the $\bf{R}, \bf{\overline R}$ representations of the gauge group, produces a coupling
\be
\Delta {\cal L} = 2 \mu({\bf R}) \frac{g^2}{32 \pi^2} \arg(m) F^a_{\mu \nu} \tF^{a\mu\nu},
\ee
where $\mu({\bf R})$ is the Dynkin index of the representation. For a $U(1)$ gauge theory, $\mu({\bf R})$ is simply $q^2$ with $q$ the quantized charge.
Applying this general formula to our model, we have the axion couplings as
\beq 
&&\frac{\alpha_h}{8\pi F_0} \left[2\mu_h(Q_{1a})(a_1 + a_2) + 2 \mu_h(Q_{1c}) a_1 \right] H \tilde{H} + \frac{\alpha_s}{8 \pi F_0}\left[2\mu_c(Q_{1b}) (a_1+a_2) \right] G \tilde{G} \nonumber \\
 && +\frac{\alpha_{\rm em}}{4 \pi F_0}  D_h(Q_{1a}) a_1 F \tilde{F} \nonumber \\
 &=& \frac{\alpha_h}{8\pi F_0} \left[2M(a_1+a_2) +  a_1\right] H \tilde{H} + \frac{\alpha_s}{8 \pi F_0 } \left(a_1+a_2\right)G \tilde{G}+ \frac{\alpha_{\rm em}}{4 \pi F_0} (M^2-1) a_1 F \tilde{F}, \nonumber \\
 \label{eq:align}
\eeq
where in the first two lines, $\mu$'s ($D$) are the Dynkin indices (dimension) of the corresponding $Q$'s in the brackets. 

To map this to the earlier more schematic discussion, note that $a_1 + a_2$ plays the role of $b$ and $a_1$ plays the role of $a$ above. After integrating out the heavy combination $(2M+1)a_1 + 2Ma_2$, the QCD axion $a$ is the light linear combination
\begin{align}
a &= \frac{F_0}{F_a} \left[2 M a_1 - (2M + 1) a_2\right], \quad {\rm where} \\
F_a &= \sqrt{8 M^2 + 4 M + 1} F_0
\end{align}
sets the period of the light field, $a \cong a + 2\pi F_a$. In the large $M$ limit, the QCD axion is approximately $a_1 - a_2$. Its period $F_a$ is larger than $F_0$ because the light field winds multiple times around the two-axion space; see e.g.~Figure 1 of Ref.~\cite{Choi:2014rja} for an illustration. The coupling to QCD determines the effective decay constant $f_a$ of the light field; in this case we find $f_a = F_a$, i.e.~the number $N$ in equation (\ref{eq:QCDcouplingdef}) is 1. 
The QCD axion coupling to photons is enhanced by
\beq
r= 4M(M^2-1),
\eeq
which could be of order 100 - 1000 for moderately large $M$ ($ 3< M < 10$).

A few comments on model building are in order:
\begin{itemize}
\item In general, from the first line of Eq. \ref{eq:align}, 
\beq
r = \frac{ D_h(Q_{1a}) \mu_h(Q_{1a}) }{\mu_c(Q_{1b}) \mu_h(Q_{1c})} 
\eeq
To maximize $r$, $Q_{1b}$ and $Q_{1c}$ should be in the fundamental
representations of $SU(3)_c$ and $SU(M)_h$ respectively, with the
smallest possible Dynkin index $\mu = \frac{1}{2}$. $D$ and $\mu$ of
the symmetric $k$-index
representation of $SU(M)$ are ${{M+k-1}\choose{k}}$ and
$\frac12{{M+k}\choose{k-1}}$
respectively
with the positive integer $k \geq 1$. Thus 
\beq
r =
2
{{M+k-1}\choose{k}}
{{M+k}\choose{k-1}}
, \quad k=1,2,3,\cdots
\eeq
If $Q_{1a}$ and $Q_{2a}$ both transform in the fundamental
representation of $SU(M)_h$, $r \sim M$.  If we ignore the constraint
on the rank of the representation of the fermions, higher
representations lead to higher power of enhancement in axion-photon
coupling. Yet high dimensional representations are also more severely
constrained by the Planckian Landau pole requirement and pose more
challenges to be UV completed in string theory. 
\item We don't consider models with KSVZ fermions transforming under
  both $SU(M)_h$ and $SU(3)_c$. Let's take a look at a simple model
  containing such fermions. Consider a vector-like fermion pair
  $Q_{1}({\widetilde Q}_{1})$, charged under both $SU(M)_h$ and
  $SU(3)_c$, coupling to the heavy axion $a_1$. To realize the
  alignment mechanism, we need another vector-like fermion pair
  $Q_{2}$ ($\tilde{Q}_{2}$), transforming under $SU(M)_h$ and coupling
  to the second axion $a_2$, which is mostly the QCD axion. These
  fermions will generate axion couplings to the hidden and QCD gluons
  as 
\beq
\left[\mu_h(Q_{1}) D_c(Q_{1}) a_1+ \mu_h(Q_{2}) D_c(Q_{2})a_2 \right] H \tilde{H} + \mu_c(Q_{1}) D_h(Q_{1}) a_1   G \tilde{G} , \nonumber
\eeq 
where we only write down the group theoretical factors explicitly. Again we assume these fermions are weak singlets for simplicity. Integrating out the heavy fermions and the heavy axion $a_1$, we have the QCD axion-gluon coupling as 
\beq
\frac{\mu_c(Q_{1}) D_h(Q_{1}) \mu_h(Q_{2}) D_c(Q_{2})}{\mu_h(Q_{1}) D_c(Q_{1})} a_2 G \tilde{G} \sim \frac{M}{3} \mu_h(Q_{2}) D_c(Q_{2}) a_2 G \tilde{G},
\eeq
Compared to Eq.~\ref{eq:align}, this model does not lead to a
parametric enhancement of $f_a$ relative to the fundamental period $F_0$ in the UV, as we want. 
\item In general, we do not require the heavy fermions to decay. They could be (meta)stable and phenomenologically viable as long as the inflation scale is below the confinement scale of the hidden gauge group.
In the specific model,  $Q_{1a}$ could decay through high-dimensional
operator $\frac{Q_{1a}Q_{1c}Q_{1c} e^c \phi_1^3}{M_{\rm pl}^5}$ with
$e^c$ the right-handed lepton in the standard model, which respects
the PQ symmetries. Provided $Q_{1c}$ is lighter than $Q_{1a}$ and
$m_{Q_{1a}} \gtrsim 10^{15}$ GeV, the lifetime of $Q_{1a}$ is shorter
than $\sim10^{-2}\ \mathrm{s}$, so that $Q_{1a}$ decays before BBN. The
other fermions are stable on the
cosmological scale. Yet the model could be modified slightly to make
the rest of the fermions decay as well. For instance, changing the $U(1)_Y$
assignments of $Q_{1b}$ and $Q_{2b}$ to $-1/3$ or $2/3$ allow
dimension-four operators that mix these heavy quarks with the standard
model quarks and induce them to decay. These changes won't affect the
axion--photon coupling enhancement significantly. 
\item The domain wall number is $1$ in our model because $f_a = F_a$. Thus the model does not have a potential domain wall problem. In general, the domain wall problem could be solved by introducing a small explicit soft breaking of PQ symmetry that doesn't spoil the axion quality~\cite{Sikivie:1982qv}.
\item Although gauge coupling unification is a nice and desirable feature in general, we will not use it a necessary requirement to restrict the representations and heavy fermion masses in our discussions. 
\end{itemize}

\subsection{Landau Pole Constraint}
\begin{table}[h!]
\centering
	\begin{tabular}{| c  | c | c | c|}
	\hline
	$M$ & $m_Q$ (GeV) &  $f_a$ (GeV) & $r=4M(M^2-1)$ \\ \hline
	$3$ & $1.3 \times 10^{4}$ & $6.0 \times 10^{5}$ & 96 \\ 
	$4$ & $5.5 \times 10^{10}$ & $3.3 \times 10^{12}$ & 240 \\
	$5$ & $3.6 \times 10^{13}$ & $2.7 \times 10^{15}$ & 480 \\
	$6$ & $1.0 \times 10^{15}$ & $9.1 \times 10^{16}$ & 840\\
	$7$ & $7.3 \times 10^{15}$ & $7.5 \times 10^{17}$ & 1344\\
	$8$ & $2.6 \times 10^{16}$ & $3.0 \times 10^{18}$ & 2016 \\
	$9$ & $5.9 \times 10^{16}$ & $7.7 \times 10^{18}$ & 2880 \\
	$10$ & $1.0 \times 10^{17}$ & $1.5 \times 10^{19}$ & 3960 \\
       \hline
\end{tabular}
\caption{For a given choice of $M$ for an $SU(M)_h$ gauge group, the smallest $m_Q$ and corresponding decay constant $f_a$ for which we do not hit a Landau pole below $10^{18}$ GeV,  and the corresponding enhancement factor $r$ of the axion--photon coupling.  We assume all the heavy KSVZ fermions have the same mass $m_Q = y F_0$ where we have chosen $y = 0.2$ as a reference value.} \label{table:N}
\end{table}

The additional KSVZ vector-like fermions modify the RG running of the SM gauge couplings. In the model in the previous section, all non-Abelian gauge groups are asymptotically free. Yet the charged fermions will accelerate the running of $U(1)_Y$ towards large values and lower its Landau pole. We solve the two-loop RG equations numerically to compute the running of the gauge couplings. The two-loop RG equations could be found in Appendix~\ref{appA}. For simplicity we will assume that all the $Q$'s have the same mass and set the hidden gauge coupling to be 1 at the scale of $m_Q$. If we demand the Landau poles of $U(1)_Y$ to be above the Planck scale ($\gtrsim 10^{18}$ GeV), the minimum allowed vectorlike fermion mass as a function of the degree of the hidden gauge group, $M$, is shown in Table~\ref{table:N}. Notice that $m_Q = y F_0$ which is below the effective decay constant $f_a$.

\section{Scenario II: Confinement Tower}
\label{sec:model3}

Now let's extend the bi-axion alignment model to a multi-axion
alignment model. We will demonstrate that just as we use KSVZ to UV
complete the KNP alignment, we could apply KSVZ to build up a
clockwork, which offers a simple way to realize the clockwork
  structure~\cite{Choi:2015fiu}. 

Consider $n-1$ hidden gauge groups $(SU(M)_h)^{n-1}$ confining at
scales $\Lambda_1, \Lambda_2, \cdots, \Lambda_{n-1} \gg \Lambda_{\rm
QCD}$. There are $n$ PQ fields $\phi_i, i=1,2,\cdots, n$, breaking at
a common high energy scale $F_0 > \Lambda$ (we choose a common $F_0$ 
for simplicity), resulting in $n$ axion fields $a_i$'s.
The Lagrangian can be schematically written as
\beq
{\cal L} &=& 
\frac{1}{8 \pi F_0} \left[ 
  \left( \sum_{i=1}^{n-1}\left(Ma_{i}+a_{i+1} \right) 
  \alpha_{i} H_{i} \tilde{H}_{i} \right)  
+a_1  \alpha_s G \tilde{G} + a_n \alpha_{\rm em} F\tilde{F}\right],
\eeq
where $H_i$ is the field strength of the $i$th $SU(M)_h$. The potential for the axions is
\beq
V &=& 
 \Lambda_{\rm QCD}^4 \cos\left(\frac{a_1}{F_0}\right) 
+\sum_{i=1}^{n-1} \Lambda_{i}^4 \cos\left(\frac{Ma_{i} +a_{i+1}}{F_0}
\right) 
\,.
\eeq
Integrating out the heavy axions $a_1, a_2, \cdots, a_{n-1}$ could be
done by setting the arguments $Ma_{i} + a_{i+1} =0, i=1, \ldots n-1$. Again in a concrete model, there could be order-one coefficients in front of the axion fields. This leads to 
\beq
a_1 \approx \frac{a_n}{M^{n-1}},
\label{eq:tower}
\eeq
where we ignore $(-1)^{n-1}$. 
Then the effective Lagrangian of the lightest axion $a_n$, which is identified as the QCD axion, to be
\beq
{\cal L} =\frac{a_n}{8 \pi M^{n-1} F_0}\left[  \alpha_s G \tilde{G} + M^{n-1} \alpha_{\rm em} F\tilde{F} \right].
\eeq
Thus the effective decay constant is $f_a= M^{n-1} F_0$. The coefficient of the axion-photon coupling is enhanced by $r = M^{n-1}$, which could be arbitrarily large in principle.

This mechanism could be realized in a KSVZ model easily as well. The particle content of one particular model is shown in Table~\ref{table:content2}. In this model, 
\beq
r= (2M)^{n-1}. 
\eeq
For $M=3$, $n=5$ gives $r = 1296$ and $n=9$ gives $r=1.6 \times 10^6$. There is no Landau pole issue in this model since only one set of vector-like fermions is charged under $U(1)_Y$. The vector-like fermions and the radial modes of the PQ fields have masses of order $F_0$. The heavy axions have masses of order $\Lambda^2/F_0$. Depending on $M$ and $n$, these particles could be relatively light and may be even close to the TeV scale to be probed directly at the LHC or future colliders. We will leave this model-dependent phenomenology for future work.

The low energy spectrum of axions in our model is very similar to that
of the clockwork based on many scalars with a particular type of
quartic interaction in Ref.~\cite{Farina:2016tgd}. One explicit way to
see the similarity is that the mass matrices for the axions in both
models take the same tridiagonal form. Yet in our model, the clockwork
is based on confining gauge groups and fermions with small
representations. This might be more easily realized in the UV than a
set of scalars with $1/3^n$ charge assignments. 

The ``confinement tower'' construction we present here is very similar to a model presented in Section III of \cite{Choi:2015fiu}, which differs in representation choice and in being supersymmetric.
Our scenario also bears some similarity to models recently proposed in
Ref.~\cite{Coy:2017yex}, in which the axion with a small $F_0$ and a
large $f_a$ arises as a Goldstone from a set of confining gauge groups
as well. Our model is KSVZ type with heavy fermions' mass above the
confining scale while models in Ref.~\cite{Coy:2017yex} rely on the
condensation of fermions. We have several elementary axions while
their models involve mostly composite axions.

\begin{table}[h!]
\centering
	\begin{tabular}{| c  | c c c c c  c  |}
	\hline
	 & $SU(M)_{h;1}$ &$SU(N)_{h;2}$& $\cdots$ & $SU(N)_{h;n-1}$ & $SU(3)_C$ & $U(1)_Y$	 \\ \hline \hline
	 $\phi_1$ & 1  & 1 & 1 &1 & 1 & 0 \\ 
	 $Q_{1a} (\tilde{Q}_{1a}) $ & Adj & 1 &1 &1 &1  &0 \\ 
	 $Q_{1b} (\tilde{Q}_{1b})$ & 1 &1 &1 &1 & 3 ($\overline{3}$) & 0 \\ \hline
	  $\phi_2$ & 1  & 1 & 1& 1&  1 & 0 \\ 
         $Q_{2a} (\tilde{Q}_{2a}) $ & $M$ ($\overline{M}$) & 1&  1 & 1& 1&  0   \\  
	 $Q_{2b} (\tilde{Q}_{2b})$ & 1 & Adj & 1 &1  & 1 & 0 \\ \hline
	 	  $\phi_3$ & 1  & 1 & 1& 1&  1 & 0 \\ 
         $Q_{3a} (\tilde{Q}_{3a}) $ &1  & $M$ ($\overline{M}$)&  1 & 1& 1&  0   \\  
	 $Q_{3b} (\tilde{Q}_{3b})$ & 1 & 1& Adj &1  & 1 & 0 \\ \hline
	 $\cdots$ &  $\cdots$ &  $\cdots$ & $\cdots$ &  $\cdots$ &  $\cdots$ &  $\cdots$ \\ \hline
	 	  $\phi_n$ & 1  & 1 & 1& 1&  1 & 0 \\  
         $Q_{na} (\tilde{Q}_{na}) $ &1 & 1&  1 & $M$ ($\overline{M}$)&  1 &  0   \\  
	 $Q_{nb} (\tilde{Q}_{nb})$ & 1 & 1 & 1 &1  & 1 & $1$ ($-1$) \\  \hline
	 \end{tabular}
\caption{Particle content of a confinement tower model. } \label{table:content2}
\end{table}

\subsection{Axion Quality}

It is known that a global continuous symmetry is not respected by quantum gravity~\cite{Abbott:1989jw, Coleman:1989zu, Kamionkowski:1992mf, Kallosh:1995hi, Banks:2010zn, Alonso:2017avz} and we generally expect high dimensional operators suppressed by Planck scale that break the global PQ symmetry to appear. These operators tend to generate too large a strong CP phase and ruin the PQ mechanism. One possible way to suppress the dangerous operators, though baroque, is to invoke a discrete symmetry $\mathbb{Z}_N$ with $N \gg 1$ to suppress up to dimension 10 operators. While the alignment model based on a single confining gauge group in Sec.~\ref{sec:model2} is subject to the same issue as the standard KSVZ model, the axion quality in the confinement tower scenario could be significantly improved. This is due to the exponential enhancement of the effective decay constant $f_a$ over the fundamental period $F_0$ in the UV. For example, consider the dimension-five operator $c_1 \phi_n^5/M_{\rm pl}$ which explicitly breaks the PQ symmetry. It will contribute to the axion potential 
\beq
\delta V = \frac{|c_1| F_0^5}{M_{\rm pl}} \cos \left(\frac{5a_n}{F_0} - \alpha \right) \quad {\rm with} \quad \alpha= {\rm arg}[c_1],
\eeq
which shifts the strong CP phase by 
\beq
\delta \theta \approx \alpha \frac{F_0}{5 f_a} \approx \frac{\alpha}{5 r}.
\eeq
For a generic complex coefficient $c_1$, $\alpha \sim {\cal O}(1)$.
The larger the enhancement factor $r$ is, the smaller the shift in
$\theta$ is. If $r > 10^{10}$, the axion is immune to the PQ breaking
high-dimensional operators. For smaller $r$, the axion quality problem
is alleviated such that we only need to introduce some much smaller
discrete symmetry to protect the axion.
This argument also applies to the original scalar clockwork scenario~\cite{Higaki:2016yqk}.

\section{Scenario III: Kinetic Mixing of Multiple Axions}
\label{sec:model4}

A set of axions can, in general, kinetically mix with each other \cite{Babu:1994id}. This idea has found various applications in phenomenology; for instance, it has been used to produce a model in which a 7 keV dark matter axion decays to X-rays by mixing with a lighter QCD axion \cite{Higaki:2014qua}. Here we will demonstrate that the QCD axion can potentially obtain larger couplings to photons by mixing with a lighter axion field.

To this end, we consider the following model: we have a QCD axion field $a(x)$, a new confining gauge group with field strength $H_{\mu \nu}$, and an axion $b(x)$ that obtains a mass when $H$ confines. We are interested in the limit $m_a \gg m_b$. We also assume that $b$ couples to photons, and that $a$ and $b$ kinetically mix. The Lagrangian is
\begin{align}
{\cal L} &= \frac{1}{2} \partial_\mu a \, \partial^\mu a + \frac{1}{2} \partial_\mu b \, \partial^\mu b + \epsilon\, \partial_\mu a\, \partial^\mu b + c_b \frac{\alpha}{8 \pi} \frac{b}{F_b} F_{\mu \nu} \tF^{\mu \nu} + \frac{\alpha_H}{8 \pi}   \frac{b}{F_b} H_{\mu \nu} \tH^{\mu \nu} + \frac{\alpha_s}{8 \pi}   \frac{a}{F_a} G_{\mu \nu} \tG^{\mu \nu} \nonumber \\
&\to \frac{1}{2} \partial_\mu a\, \partial^\mu a + \frac{1}{2} \partial_\mu b\, \partial^\mu b + \epsilon\, \partial_\mu a \,\partial^\mu b + c_b \frac{\alpha}{8 \pi}  \frac{b}{F_b} F_{\mu \nu} \tF^{\mu \nu} - V_G(a) - V_H(b). 
\end{align}
Because $\epsilon$ is a continuous quantity and $a F \tF$ couplings are quantized, it might at first seem that the kinetic mixing cannot induce an effective coupling of the heavy axion $a$ to photons. However, if we diagonalize the kinetic and mass terms we see that the independently propagating axion fields are misaligned with the basis in which the axions have well-defined periodicity, which allows for more general couplings.

The physics is easiest to understand by first imagining the limit in which $b$ remains massless. In this case, following a prescription familiar from the physics of dark photons \cite{Holdom:1985ag}, we eliminate the kinetic mixing with a field redefinition, $b \to b - \epsilon a$, after which we must divide $a$ by $\sqrt{1 - \epsilon^2}$ to canonically normalize it. This has the effect of leaving terms like $a G \tG$ unchanged, so the heavy field has no admixture of $b$. However, the coupling of $b$ to photons now induces a coupling of the redefined $a$ to photons:
\be
{\cal L}_{\rm diag} \supset \left(\frac{c_b \epsilon F_a}{F_b}  + {\cal O}(\epsilon^2)\right) \frac{\alpha}{8\pi} \frac{a}{F_a} F_{\mu \nu} \tF^{\mu \nu}.
\ee
The couplings of the light field $b$ remain quantized after this field redefinition (much as a massless photon always couples to a well-defined conserved current), but the heavier field $a$ acquires a new coupling of order $\epsilon F_a/F_b$. In particular, if the kinetic mixing is large and if $F_b \ll F_a$, the mixing contribution to $g_{a\gamma\gamma}$ can overwhelm more direct contributions.

We can now reintroduce the mass of $b$, which will give subleading
corrections to the QCD axion's couplings of order $m_b^2/m_a^2$
relative to the corrections considered above. It will also allow the
lighter field $b$ to obtain non-quantized couplings by mixing with
$a$, but these will be suppressed not only by $\epsilon$ but also by
$m_b^2/m_a^2$. For this reason, models where the QCD axion mixes with
{\em heavier} axions are less effective at modifying its couplings to
photons. In that case, keeping $a$ as the QCD axion, we have $m_b > m_a$ and the photon coupling behaves as $\epsilon (F_a/F_b) (m_a^2/m_b^2)$. Hence, it would be difficult to enhance the photon coupling.

\subsection{Realizing Large Mixing}

In renormalizable KSVZ or DFSZ-like axion models, we expect kinetic
mixing effects to generally be small. However, in scenarios where
axions come from higher dimensional gauge fields, the topology of
internal dimensions can lead to sizable mixing effects. We may have,
for instance,  a supersymmetric completion containing a variety of
dimensionless moduli fields $T_i = \tau_i + \frac{\iu}{2\pi} \theta_i$
where $\theta_i \cong \theta_i + 2\pi$ are periodic axion fields. The
perturbative K\"ahler potential $K(T_i + T^\dagger_i)$ depends on the
real moduli but not the axions. Axion kinetic terms arise from
derivatives of the K\"ahler potential; if the K\"ahler potential
depends in a sufficiently general way on the $\tau_i$, these kinetic
terms may be highly mixed. A variety of examples are discussed in
\cite{Cicoli:2012sz}. The context is the Type IIB string, where the
fields $\tau_i$ parametrize the volumes of 4-dimensional cycles within
a 6d compactification. The K\"ahler potential depends on the volume of
the internal dimensions in string units, 
\be
K = - 2 M_{\rm Pl}^2 \log({\cal V}),
\ee
where ${\cal V}$ is a function of the $\tau_i$. Let us extract a
simplified version of one illustrative example from \S4.4
in~\cite{Cicoli:2012sz}. We may have
\be
{\cal V} = {\cal V}_0 - \alpha_1 \tau_1^{3/2} - \beta_1 \tau_1^{1/2} \tau_2 - \beta_2 \tau_1 \tau_2^{1/2} - \alpha_2 \tau_2^{3/2},
\ee
where ${\cal V}_0$ is a large overall volume stabilized in a way that effectively decouples from the fields $\tau_{1,2}$, and the $\alpha_i$ and $\beta_i$ are some order-one coefficients, calculable from topological data (intersection numbers).

In this case, one readily calculates that the kinetic terms for the axions have the form:
\be
\frac{1}{2} F_1^2 \partial_\mu \theta_1 \partial^\mu \theta_1 + \epsilon F_1 F_2 \partial_\mu \theta_2 \partial^\mu \theta_2 + \frac{1}{2} F_2^2 \partial_\mu \theta_2 \partial^\mu \theta_2,
\ee
with
\begin{align}
F_1^2 &\approx \frac{M_{\rm Pl}^2}{{\cal V}_0} \frac{3 \alpha_1 \langle \tau_1\rangle - \beta_1 \langle \tau_2 \rangle}{4 \langle \tau_1\rangle^{3/2}} + {\cal O}(M_{\rm Pl}^2/{\cal V}_0^2), \nonumber \\
F_2^2 &\approx \frac{M_{\rm Pl}^2}{{\cal V}_0} \frac{3 \alpha_2 \langle \tau_2\rangle - \beta_2 \langle \tau_1 \rangle}{4 \langle \tau_2\rangle^{3/2}} + {\cal O}(M_{\rm Pl}^2/{\cal V}_0^2), \nonumber \\
\epsilon &\approx \frac{2 \beta_1  \langle \tau_1\rangle^{1/4}  \langle \tau_2\rangle^{3/4} + 2 \beta_2  \langle \tau_1\rangle^{3/4}  \langle \tau_2\rangle^{1/4}}{(3 \alpha_1 \langle \tau_1\rangle - \beta_1 \langle \tau_2 \rangle)^{1/2} (3 \alpha_2 \langle \tau_2\rangle - \beta_2 \langle \tau_1 \rangle)^{1/2}} + {\cal O}(1/{\cal V}_0).
\end{align}
In this example we see that:
\begin{itemize}
\item If ${\cal V}_0 \gg 1$, all decay constants are well below the Planck scale: the prefactor is set by the string scale $M_{\rm string} \sim M_{\rm Pl}/\sqrt{{\cal V}_0}$.
\item If $\langle \tau_1 \rangle \sim \langle \tau_2 \rangle$, then the two decay constants are parametrically the same size and their mixing $\epsilon$ is ${\cal O}(1)$.
\item The example from \cite{Cicoli:2012sz} has $\alpha_1, \alpha_2, \beta_1 > 0$ and $\beta_2 < 0$. In this case, we can avoid ghosts if $\langle \tau_1 \rangle \gg \langle \tau_2 \rangle$ but not vice versa. The hierarchy of decay constants in this limit is $F_2/F_1 \sim \langle \tau_1/\tau_2 \rangle^{3/4} \gg 1$ and the kinetic mixing is suppressed by
\be
\epsilon \sim \langle \tau_2/\tau_1 \rangle^{1/4} \sim (F_1 / F_2)^{1/3}.   \label{eq:kineticscaling}
\ee
If the axion $\theta_2$ obtains a much larger mass than $\theta_1$, then the couplings of $\theta_2$ to gauge fields to which $\theta_1$ couples with order-one strength can be enhanced by the large ratio $\epsilon F_2/F_1 \sim (F_2/F_1)^{2/3}$.
\end{itemize}
It is not a stretch to believe that axions can have large kinetic mixing in string theory; the structure of the K\"ahler potential makes it generic for general enough topology. The ingredient that may be somewhat more tricky to realize is a large hierarchy $\langle \tau_1 \rangle \gg \langle \tau_2 \rangle$ between the volumes of different cycles. For now, we simply observe that we have transmuted a problem of obtaining large axion couplings into a problem of obtaining geometric hierarchies from moduli stabilization. There is a rich literature on moduli stabilization that makes it plausible that such hierarchies can be realized.

In this discussion we have focused on kinetic mixing between just two axions. In theories with a large number of axions, more dramatic effects may be possible. A recent general analysis of kinetic and St\"uckelberg mixings for multiple axions, including effects on the field range and couplings, appeared in \cite{Shiu:2015uva, Shiu:2015xda}. The phenomenon of kinetic alignment can arise, with a randomly chosen kinetic matrix having a very large eigenvalue compared to the typical size of the other eigenvalues \cite{Bachlechner:2014hsa, Bachlechner:2014gfa, Bachlechner:2017zpb, Bachlechner:2017hsj}. This is a distinct phenomenon from KNP alignment, which relies on special structure in the charge assignments of the instantons giving rise to dominant contributions to the axion potential. Kinetic alignment has been studied in the inflationary context, where it provides an interesting test case for arguments for or against the ability of quantum gravity to accommodate super-Planckian field ranges \cite{Bachlechner:2014gfa, Rudelius:2014wla, Bachlechner:2015qja, Montero:2015ofa, Brown:2015lia, Junghans:2015hba, Heidenreich:2015wga}. It has not yet been applied to more general axion phenomenology, where new mechanisms for sub-Planckian field ranges are already of interest. We will leave consideration of many-axion kinetic mixing for future work.

\section{Results and Conclusions}

\begin{figure}[th]
  \centering
  \includegraphics[width=0.95\textwidth]{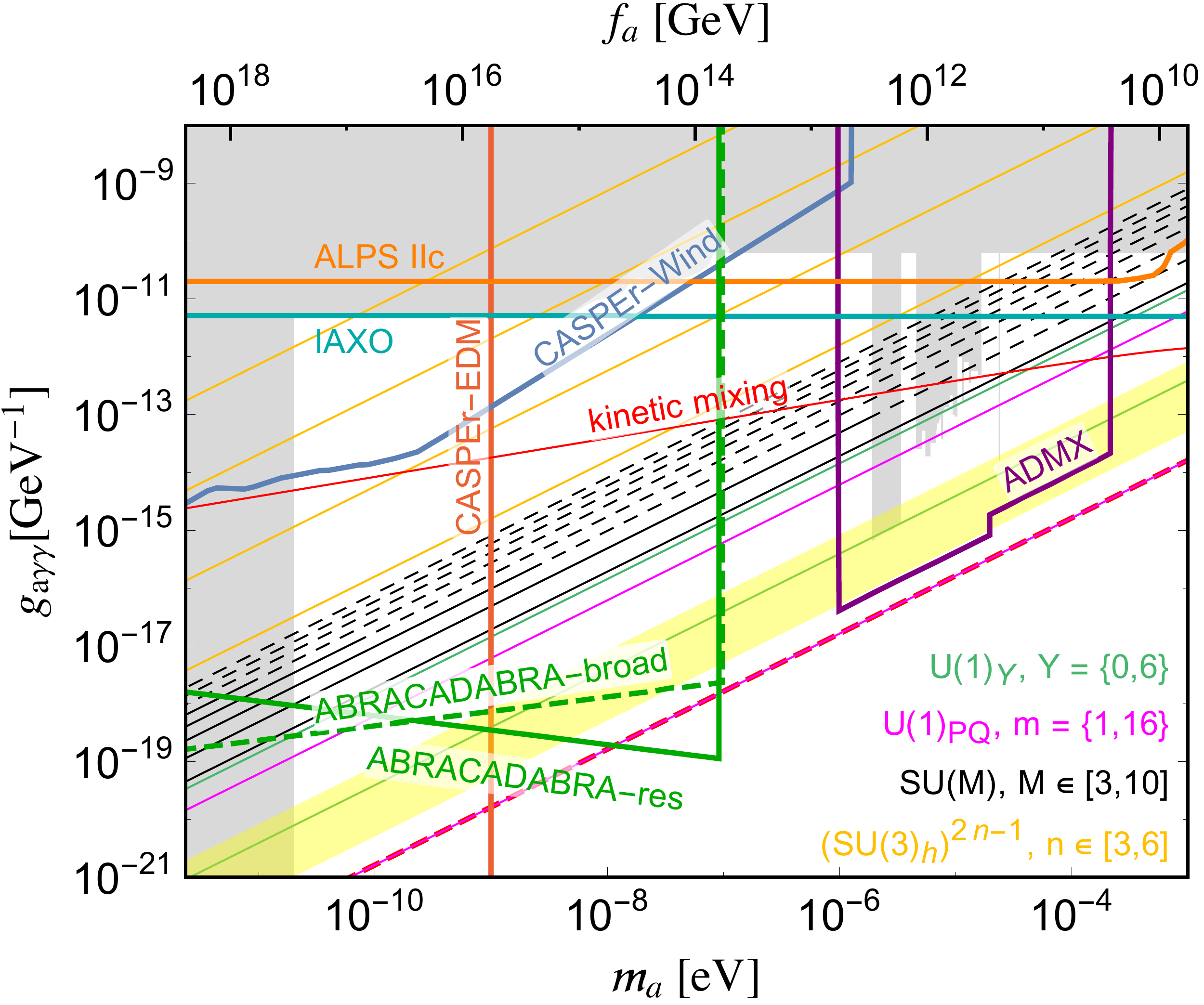}
  \caption{Parameter space of the axion coupling
    with photons. The gray regions are existing limits. Thin slanted lines
    represent the predictions in models of
    sections~\ref{sec:model2} (black), \ref{sec:model3} (amber), and
    \ref{sec:model4} (red) and
    appendix \ref{sec:model1} (green, magenta) are shown explicitly.
    Regions bounded by colored lines show sensitivity of upcoming
    experiments, ALPS II (orange), IAXO (teal), CASPEr-Wind (blue),
    CASPEr-EDM (red), ABRACADABRA (green) and ADMX (purple). Further
    details for the plot can be found in the text.
  \label{fig:photofriend}
}
\end{figure}

In figure~\ref{fig:photofriend} we show the parameter space of the
models in the $m_a$--$g_{a\gamma\gamma}$ plane.
The current constraints are shown as gray shaded regions, which 
arise from evolution of
horizontal branch stars~\cite{Cadamuro:2011fd}, from the
CAST helioscope~\cite{Anastassopoulos:2017ftl}, and microwave
cavities such as
ADMX~\cite{DePanfilis:1987dk,Wuensch:1989sa,Hagmann:1990tj,Asztalos:2009yp,Brubaker:2016ktl}.
The strongest
constraint on light
axions arises from non-observation
of axions from SN1987A~\cite{Payez:2014xsa} 
and from
conversion of X-ray photons to axions in cluster magnetic fields
\cite{Berg:2016ese,Marsh:2017yvc,Conlon:2017qcw}.
Observation of
black hole spins disfavors a range of axion masses which would lead to
superradiance of the black hole~\cite{Arvanitaki:2014wva}. There are
additional constraints from
observations of the  gamma-ray
spectra
by
HESS~\cite{Abramowski:2013oea} 
and Fermi-LAT~\cite{TheFermi-LAT:2016zue}.

A number of future experiments will cover the unexplored parameter
space for the QCD axion and axion-like particles. ADMX will extend its
reach to axion dark matter for a wider mass range and to higher
sensitivities~\cite{Shokair:2014rna}.
The ``light shining through walls'' experiment ALPS
II~\cite{Bahre:2013ywa},
and the helioscope IAXO~\cite{Vogel:2013bta} will be sensitive to
large axion photon
couplings. 
There are new experimental proposals such as
CASPEr-EDM and
CASPEr-Wind~\cite{Graham:2013gfa,Budker:2013hfa,Garcon:2017ixh}
which use NMR, and 
ABRACADABRA~\cite{Kahn:2016aff} which is a broadband/resonant search
for magnetic fluxes induced by axions in a background magnetic field.
These experiments can probe light axions down to the
QCD line. We show projected sensitivities from future experiments as
regions bounded
by colored solid lines in figure~\ref{fig:photofriend}.

Overlaid on the experimental reach, we show
the possible values of $g_{a\gamma\gamma}$ obtained in the
models in sections~\ref{sec:model2}, \ref{sec:model3} and
\ref{sec:model4}, and in appendix~\ref{sec:model1}. 
The traditional axion-photon coupling band (as discussed
in~\cite{Kim:1998va}) is
shown as the pale yellow shaded region.
With black lines we show the possible enhancement of
$g_{a\gamma\gamma}$ for the model with a single confining gauge
group presented in section~\ref{sec:model2}. In this case we have
estimated the minimum $f_a$ required for the hypercharge Landau pole
to be below the Planck scale, and we show $f_a$ values smaller than
this value as
dashed lines. We note however that this is not a strict constraint on
the model space, as a lower Landau pole or a different value of the
Yukawa coupling is possible.
Simple extensions of the KSVZ model with large hypercharges or large
PQ charges are also shown, subject to the constraints discussed
in appendix~\ref{sec:model1}; we see that they can only get moderate
enhancements relative to the more traditional KSVZ models.
Including the
effect of the confining tower clockwork model in section~\ref{sec:model3}
lets us cover the entire parameter space, similar to
Ref.~\cite{Farina:2016tgd}. We have shown (in amber) a
particular realization with varying number of copies of the confining
group, which is chosen to be $SU(3)$.
We also show an example of the enhancement we can obtain by
kinetically mixing the QCD axion with another lighter axion. For the
lighter axion we chose the coupling to photons at the limit with the
mass to be $10^{-13}\ \mathrm{eV}$, 
i.e.~$g_{a\gamma\gamma}=5.34\times10^{-12}\ \mathrm{GeV}^{-1}$.
For concreteness we assume that (before the field redefinition to remove mixing) the light axion coupling to the
photon is $g_{a\gamma\gamma} = \alpha_{\rm em}/{(2\pi F_1)}$, and the QCD
axion coupling is $g_{a\gamma\gamma} = -1.92\alpha_{\rm em}/(2\pi F_2)$,
with $F_2=f_a$.
Then, the
maximum enhancement as in equation~\ref{eq:kineticscaling} is represented by the
red line in figure~\ref{fig:photofriend}.  We note that mixing with a
lighter state can lead to significant
deviations from the quantized discretuum of
$g_{a\gamma\gamma}$. Finally, without tuned contributions from
multiple representations or the kinetic mixing, the smallest
$g_{a\gamma\gamma}$ that can be obtained simply is expected from
$E/N$=2, and is shown as a red dashed line.

The QCD axion remains a very well-motivated dark matter candidate,
with exciting upcoming experiments searching for its couplings to
photons. We have shown that minimal extensions to the simplest models
can lead to a large enhancement of axion--photon couplings, making most
of the open parameter space a promising target to look for QCD axions.

\acknowledgments{MR and LTW thank the 29th Recontres de Blois for providing a congenial environment to begin discussing this topic. A portion of this work was carried out at the Aspen Center for Physics, which is supported by the National Science Foundation grant PHY-1607611. PA is supported by the NSF grants PHY-0855591 and PHY-1216270. 
JF is supported by the DOE grant
DE-SC-0010010.  MR is supported in part by the DOE grant DE-SC0013607
and the NASA grant NNX16AI12G. LTW is supported by the DOE grant DE-SC0013642.}

\appendix

\section{Mass Mixing and Non-Quantized Couplings}
\label{massmix}

In the introduction we asserted that couplings of the form $a F \tF$ are quantized, but then displayed an axion--pion mixing effect (\ref{eq:axionpionmixing}) that appeared to violate the quantization. For clarity, we will explain this apparent contradiction. The short answer is that such couplings are part of a larger set of couplings summing up to a periodic function of the axion, and as such are always accompanied by a factor of $m_a^2$ as a spurion for the breaking of the continuous shift symmetry. In particular, it is only because $\Lambda_{\rm QCD}^4 \sim m_a^2 F_a^2$ that this effect is sizable. More general mass mixing will, as a rule, lead to negligible non-quantized effects.

Consider the following very schematic toy model for axion--meson
mixing, which simplifies the situation in real QCD by considering only a single meson, which we denote $\pi^0$ (though in the one flavor case it behaves more like the $\eta'$; the full theory includes several mesons that all mix). We assume this meson couples to photons through a Lagrangian
\be
{\cal L}_{\rm int} = \frac{\alpha}{8 \pi} 
\frac{\pi^0}{f} F_{\mu \nu} \tF^{\mu \nu} 
+ \Lambda^4 \cos\left(\frac{\pi^0}{f} 
+ \frac{a}{F_a}\right)+ m_q \mu^3 \cos\left(\frac{\pi^0}{f}\right),
\ee
which has the desired property that when $m_q \to 0$ there is a
massless axion field. Here $\Lambda$ is roughly the confinement scale,
and $\mu^3$ is $|\langle q {\bar q} \rangle|$. Our goal is to see,
after integrating out the $\pi^0$, what form the axion coupling to
photons has in the low-energy effective theory. This serves to
illustrate the important physics for the QCD mixing contribution to
the axion--photon coupling, without all of the details. 

First, notice that in the $m_q \to 0$ limit, we can integrate out the
$\pi^0$ and find a coupling $-\frac{\alpha}{8\pi} \frac{a}{F_a} F
\tF$, which has an integer coefficient as expected from
(\ref{eq:quantizedcoupling}). The physics at $m_q \neq 0$ is more
interesting: if we expand the potential to quadratic order and
minimize, we find
\be
\frac{\pi^0}{f} = - \frac{\Lambda^4}{\Lambda^4 + m_q \mu^3} \frac{a}{F_a} \approx - \frac{a}{F_a} + \frac{m_q \mu^3}{\Lambda^4} \frac{a}{F_a} + \cdots,
\ee
where in the last step we see that if we expand at small quark mass we
obtain an apparently small shift away from integer values of the
coefficient of $\frac{\alpha}{8\pi} \frac{a}{F_a} F \tF$ in the
effective theory. How is this consistent with the period of the field
$a$? The answer comes from keeping the full set of nonlinear
interactions: the condition $\partial V/\partial \pi^0 = 0$ requires
\be
\frac{\pi^0}{f} 
= - \arctan 
\frac{\Lambda^4 \sin \frac{a}{F_a}}
{\Lambda^4 \cos \frac{a}{F_a} + m_q \mu^3}, 
\label{eq:nonquantizedterm}
\ee
so when we integrate out the $\pi^0$ exactly we obtain a coupling of
the form $\frac{\alpha}{8\pi} g(a/F_a) F \tF$ where $g(x)$ is a
function with period $2\pi$ and is perfectly consistent with the
periodicity of the axion.

This shows that if we consider an effective theory with general couplings
\be
{\cal L} \supset \left(c_1 \frac{a}{F_a} + c_2 \frac{a^2}{F_a^2} + c_3 \frac{a^3}{F_a^3} + \ldots\right) F_{\mu \nu} \tF^{\mu \nu},
\ee
there is in general no consistency condition on individual couplings $c_i$; rather, they can correspond to the Taylor series of any periodic function, and there is no need to impose $c_1 = n \alpha/(8 \pi)$ for integer $n$. However, there is a catch: these nonperiodic effects are always proportional to the axion mass squared. The reason is that they are sensitive to the periodicity of the axion, which means they feel the breaking of the continuous axion shift symmetry to a discrete shift symmetry. Such effects always arise from instantons, which in general contribute to the axion mass. In the current context, this is manifested in the proportionality of the non-quantized coefficient in (\ref{eq:nonquantizedterm}) to $m_q \mu^3 \sim m_a^2 F_a^2$. This can be a significant effect in QCD because the same source of nonperturbative dynamics gives mass to both the pion and the axion. Other new physics at a scale $\Lambda \approx \Lambda_{\rm QCD}$ could potentially also affect the axion couplings significantly, but would tend to spoil the strong CP solution. As a result, we do not expect mass mixing to generate significant non-quantized axion--photon couplings.

\section{Two-loop RG Equations for the Model in Sec.~\ref{sec:model2}}
\label{appA}
The two-loop RG equations for the gauge couplings, $g_i$, are given by
\beq
\frac{d \alpha_i^{-1}}{dt} = - a_i -\sum_j \frac{b_{ij}}{4\pi} \alpha_j,  \quad i = 1,2,3,h
\eeq
where $\alpha_i = \frac{g_i^2}{4\pi}$ and $t=\frac{1}{2\pi} \log{\frac{\mu}{m_Z}}$. The one- and two-loop beta functions are~\cite{Machacek:1983tz} 
\beq
a_i &=& -\frac{11}{3} C_2(G_i) + \frac{4}{3} \sum_F\kappa \mu(F_i) + \frac{1}{3} \sum_S \eta \mu(S_i), \\
b_{ij} &=& -\left(\frac{34}{3} C_2(G_i)^2-\frac{20}{3} \sum_F \kappa C_2(G_i)\mu(F_i)-\frac{2}{3} \sum_S \eta C_2(G_i)\mu(S_i)\right)\delta_{ij} \nonumber \\
&&+4 \left(\sum_F \kappa C_2(F_j)  \mu(F_i)+\sum_S \kappa C_2(S_j)  \mu(S_i)\right),
\eeq
where $F$'s are fermions and $S$'s are scalars. There is no summation
over index $i$. $\kappa = 1 (1/2)$ for Dirac (Weyl) fermions and $\eta = 1 (1/2)$ for complex (real) scalars. $G_i$ denotes the $i$th gauge factor. $C_2$'s are the Casimir of a given irreducible representation. The Dynkin indices $\mu$'s include multiplicity factors. 

Below $m_Q$ but above the confinement scale of $H$, the SM particles are decoupled from the $SU(M$) gluons and their contributions to the beta functions are 
\beq
a= \left(\frac{41}{10}, - \frac{19}{6}, -7, -\frac{11}{3} M\right), 
\eeq 
and 
\begin{align}
   b &=
    \begin{bmatrix}
    \frac{199}{50} & \frac{27}{10} & \frac{44}{5} & 0 \\
   \frac{9}{10} & \frac{35}{6} & 12 & 0 \\
   \frac{11}{10} & \frac{9}{2} & -26 & 0 \\
   0 & 0 &0 & -\frac{34}{3} M^2
    \end{bmatrix}
    \end{align}
Notice that we use the GUT normalization of the hypercharge coupling, $g_1$.
 
Above $m_Q$, the contributions of the heavy vector-like fermions in Sec.~\ref{sec:model} to the beta functions are 
\beq
\delta a = \left(\frac{4}{5} (M^2-1), 0, \frac{4}{3}, \frac{8}{3} M + \frac{2}{3}\right)
\eeq
and 
\begin{align}
  \delta b &=
    \begin{bmatrix}
  \frac{36}{25} (M^2-1) & 0 & 0 & \frac{12}{5} M (M^2-1) \\
  0 & 0 & 0 & 0 \\
0 & 0 & \frac{76}{3}  & 0 \\
   \frac{12}{5}M & 0 &0 & \frac{64}{3} M^2 + \frac{13}{3} M - \frac{1}{M}
    \end{bmatrix}
    \end{align}
We set the gauge coupling of $SU(M)_h$ to be 1 at $m_Q$. We ignore the contributions of the Yukawa couplings to the running of the gauge couplings. The Yukawa couplings lead to a much more complicated formula, which we don't include here. The Yukawa couplings between the heavy fermions and the PQ fields are free parameters and could be small. They only contribute to the gauge coupling running at the two-loop order and the effect is numerically tested to be small as long as the they are $\lesssim 1$. (In particular, we have included the Standard Model top Yukawa in the RGEs, and found no change in our conclusions about Landau poles.)

\section{Vector-like Leptons with Large PQ Charge} 
\label{sec:model1}
In this appendix, we consider the possibility of a large PQ-charged state enhancing the axion--photon coupling. It is similar to the large hypercharge case in using some large charge to increase the coupling yet suffers from different phenomenological issues, which we discuss in some detail below.  
 We study a variant of the KSVZ model with vector-like fermions which carry the large PQ charge.
Consider the following matter charge assignment with a global $U(1)_{PQ}$,
\begin{align}
  \begin{array}{|c|ccc|}
  \hline
    & SU(3)_c & U(1)_Y & U(1)_{PQ} \\
    \hline
    Q &  3 & 0 & 1\\
    \tQ & \bar{3} & 0 & 1 \\
    L &  1 & 1 & m\\
    \tL & 1 & -1 & m\\
   \phi & 1 & 0 & -2 \\
   \hline 
  \end{array}
\end{align}
where $m$ is a positive integer. All the fermions are taken to be weak singlets. 

The Lagrangian consistent with these symmetries is
\begin{align}
  \mathcal{L}
  &=
  -V(\phi)
  +
  \left(
  \lambda \phi Q \tQ
  +\lambda' \frac{\phi^m}{\Lambda^{m-1}} L \tL
  +{\rm h.c.}
  \right).
\end{align}
As usual, $V(\phi)$ is chosen to give the PQ scalar a VEV.
The Goldstone can be parametrized as
\begin{align}
  \phi = F_a \E^{\iu \frac{a}{F_a}}
\end{align}
and the
axion excitation around this VEV can be written as,
\begin{align}
  \lambda \E^{\iu \frac{a}{F_a}} Q \tQ
  +\lambda' \frac{F_a^m}{\Lambda^{m-1}} \E^{\iu \frac{ma}{F_a}} L \tL
+V(\phi).
\end{align}
In this model, the effective decay constant $f_a$ is equal to the fundamental period $F_a$. Below we will only use $f_a$. 
Upon doing the chiral rotations to get rid of the phase in the mass
terms, and integrating out the heavy fields, we get,
\begin{align}
  \frac{\alpha_s}{8\pi f_a} a G^a_{\mu\nu} \widetilde{G}^{a,\mu\nu}
  +\frac{m \alpha_{\rm em} }{4\pi f_a} a F_{\mu\nu} \widetilde{F}^{\mu\nu}
\end{align}

The potential problem here is that the mass for $L,\tL$
is suppressed by $(f_a/\Lambda)^{m-1}$. We consider two examples of
UV-completing the higher dimensional operator to see how large $m$ can
be.

\subsection{A Chain of Vector-like Fermions}
A simple renormalizable UV completion of this model is to consider a
chain of interactions,
\begin{align}
  y \phi L \tN_1 + y \phi N_{m-1} \tL
  + \sum_{i=2}^{m-2}
  y\phi N_i \tN_{i+1}
  + \sum_{i=1}^{m-1}
  M N_i \tN_i
\end{align}
where the charge assignment for $N_i,\tN_i$ is
\begin{align}
  \begin{array}{|c|ccc|}
  \hline
    & SU(3)_c & U(1)_Y & U(1)_{PQ} \\
    \hline
    N_i&  1 & 1 & m-2i\\
    \tN_i & 1 & -1 & -(m-2i)\\
    \hline
  \end{array}
\end{align}
For simplicity, we have kept the mass and the Yukawa couplings of
$N_i$ the same. Integrating out the $N_i$ at the scale $M$, 
we see that we can identify $\Lambda= M/y$, and $\lambda'=y$. 
If $M$ is smaller or close
to $f$ then the solution is essentially the same as adding a particle
with a large hypercharge. The advantage of choosing $M>f$ is that  the
hypercharge Landau pole is postponed.

\subsection{Clockwork}
Another possibility is a UV completion similar to the clockwork
mechanism. We have $l$ scalar fields $\phi_i$ with the interaction
terms
\begin{align}
  \mathcal{L}
  &=
  \lambda' \phi_0 L \tL
  +\lambda \phi_l Q \tQ
  +\sum_{i=0}^{l-1}\kappa \phi_i^\dagger \phi_{i+1}^q
  +M^2 \phi_i^\dagger \phi_i
\end{align}
The state $\phi_l$ has PQ charge 2, as evidenced by its coupling to
$Q\tQ$. The scalar $\phi_0$ has a PQ charge of $2q^l \equiv 2m$.
The mass eigenstate that is light compared to $M$ is identified as
$\phi=\phi_l$. Thus, integrating out the other $\phi_i$ at scale $M$,
\begin{align}
  \mathcal{L}
  &=
  \lambda'\left(\frac{\kappa \phi}{M^2}\right)^m L \tL
  + \lambda \phi Q \tQ
\end{align}
The effective hierarchy we get in this case is $q^l$, at the cost of
introducing only $l$ new fields with hypercharge. This model reduces
to the single axion model above only if $M > f_a$, such that all
but one axion can be integrated out. For $M\ll f_a$ the model has
multiple axions clockworking.


In both the mechanisms above, we see that if we want to integrate out
heavier physics to get a single axion effective theory at the scale
$f_a$, the mass of the $L,\tL$ fermion is suppressed exponentially. 
If we impose the condition that the fermions $L,\tL$ are heavier than
about $1\,\rm{TeV}$, 
\begin{align}
  \frac{f_a}{M}
  \geq
  \left(\frac{1\, \rm{TeV}}{M}\right)^{\frac{1}{m}}
  \geq
  \left(\frac{1\, \rm{TeV}}{M_{\rm pl}}\right)^{\frac{1}{m}}
  \simeq
  10^{-\frac{16}{m}}
\end{align}
where we have chosen $y\simeq 1, \kappa\simeq M$ for simplicity.
Therefore, we see that we can only get the enhancement factor $r=2m < 32$ if we want some
hierarchy between the scales $f_a$ and $M$.

\bibliography{ref}
\bibliographystyle{jhep}
\end{document}